\documentclass[aps,pra,twocolumn,showpacs]{revtex4}
\usepackage{amssymb}
\usepackage{color}
\usepackage{amsmath}
\usepackage{graphicx}
\usepackage{bm}

\begin{document}

\title{Quantum Carnot cycle with inner friction}

\author{Sel\c{c}uk \c{C}akmak}
\email{selcuk.cakmak@samsun.edu.tr}
\affiliation{Department of Software Engineering, Samsun University, 55420 Samsun, Turkey}

\author{Ferdi Altintas}
\email{ferdialtintas@ibu.edu.tr}
\affiliation{Department of Physics, Bolu Abant Izzet Baysal University, Bolu, 14280, Turkey}

\begin{abstract}
A single driven spin is investigated as the working substance of a six-stroke irreversible quantum Carnot cycle. The role of inner friction associated with the finite-time adiabatic transformations on the cycle efficiency and the harvested work are investigated in detail. The inner friction is found to significantly reduce the work output and the cycle efficiency which can make the engine incapable to produce positive work for the too fast adiabatic transformations. The ideal Carnot efficiency is found to be reached only for the quasi-static transformations. A deviation of the cycle efficiency from the classical Carnot efficiency has been given by an efficiency lag which is directly related to the total entropy production due to the inner friction. The released heat in the relaxation processes of the cycle are associated with the entropy production and the inner friction.  The extension of the results for a scale invariant quantum working substance and the possible experimental implementation of the irreversible quantum Carnot cycle in a liquid state nuclear magnetic resonance setup are also discussed.
\end{abstract}
\pacs{05.70.Ln,07.20.Pe}

\maketitle

\section{Introduction}
Since the recognition of a three-level maser as a heat engine operating at the Carnot limit~\cite{scovil59}, the concepts of quantum thermodynamic processes and quantum heat engines have been studied extensively~\cite{quan07,quan09,kieu04,lutz14,lutz12,fialko12,zhang14,sothmann12,quan06,altintas15,harris16,robnagel16,peterson19,assis19,zou17,klatzow19,scully03,dillen09,turkpence17,hardal15,huang14,huang12,uzdin15,uzdin16,bender2000,cakmak17,thomas14,allece15,rezek06,kosloff02,feldmann06,plastina14,deffner10,francica19,camati19,rezek10,kosloff13,quan14, xiao15, Gardas15, Lekscha18, Dann20, Allahverdyan13,batalhao15,batalhao14,allah05,campisi16}. A quantum system as a working medium extracts useful work by operating between a source bath and an entropy sink through a quantum generalization of the classical thermodynamic cycle~\cite{quan07,quan09,kieu04}. Recently, different experimentally feasible quantum systems are proposed to be engineered as a nanoscale thermal engine~\cite{lutz14,lutz12,fialko12,zhang14,sothmann12,quan06,altintas15,harris16}. Moreover, quantum heat engines are implemented experimentally in several setups~\cite{robnagel16,peterson19,assis19,zou17,klatzow19}. Both quantum and thermal fluctuations are relevant for the heat engines at nanoscale. The quantum features of the working substance and the reservoirs make the nanoscale engine to exhibit several unusual features. For instance,  a quantum heat engine fueled by the coherent resources can harvest useful work even more efficient than the classical Carnot engine~\cite{scully03,dillen09,turkpence17,hardal15,huang14,huang12}. Refs.~\cite{uzdin15,uzdin16} showed that the internal quantum coherence is responsible for the nanoscale engine to output more power than the classical counterpart using the same reservoirs. An experiment performed in nuclear magnetic resonance (NMR) platform illustrated that irreversible processes can make the quantum Otto cycle more efficient than the reversible one in the case of a negative source temperature~\cite{assis19}.

The heat-to-work efficiency of a classical heat engine is maximized for reversible processes in which the irreversible entropy production is zero. The classical Carnot engine produces work through two reversible isothermal and two reversible adiabatic processes. The efficiency of the Carnot engine is the highest among all other thermodynamic cycles. Exceeding the Carnot efficiency is attributed as the violation of the second law of thermodynamics. The general extension of the Carnot cycle for quantum mechanical systems ideally consists of two quantum versions of the isothermal and adiabatic steps~\cite{quan07,bender2000}. However, the extension is not so straightforward~\cite{quan14, xiao15, Gardas15, Lekscha18, Dann20, Allahverdyan13}. The isothermal processes can be directly extended to the quantum regime. If the isothermal process is quasi-static, the quantum working medium can be always kept in thermal equilibrium with the reservoir. Then the quantum isothermal process is reversible and the thermodynamic quantities are defined unambiguously. On the other hand, for the quantum Carnot cycle to be thermodynamically reversible, two strict conditions should be employed at the adiabatic steps~\cite{quan07}. The first condition is the requirement of the working substance to be scale invariant. That is to say, all energy gaps of the working medium should be changed by the same ratios in the two adiabatic stages. For quantum systems having only a single defined energy gap, such as a qubit and a harmonic oscillator, the scale invariance is inherently justified. For coupled systems, it is shown that scale invariance can be satisfied only when the frequency and the quantum coupling between the individuals are altered simultaneously~\cite{cakmak17}. The second condition is the requirement of the equilibrium condition at the end of the adiabatic stage. It can be succeeded by replacing adiabatic stages with quasi-static quantum adiabatic steps. In an infinitely long adiabatic process, the working substance stays in equilibrium at all times as defined by the celebrated quantum adiabatic theorem. Under the two conditions, the quantum Carnot cycle is reversible. The heat-to-work efficiency equals to the classical Carnot efficiency which is independent of the details of the quantum working substance~\cite{quan07}.

In practice, it is not possible to maintain an adiabatic step a long time due to the decoherence effects. The time length of an adiabatic step should be shorter than the characteristic decoherence time scales to avoid any heat exchange with the surrounding. Therefore, the adiabatic stages of a quantum Carnot cycle should be executed in a finite-time. However, the finite-time steps are the origin of the irreversibility~\cite{thomas14,allece15,rezek06,kosloff02,feldmann06,plastina14,deffner10,francica19,camati19,rezek10,kosloff13}. In a finite-time adiabatic process, the working medium is driven to an out of equilibrium state. When the system becomes in contact to a reservoir before the isothermal process, an inevitable thermalization with the bath occurs. In addition to two isothermal and two adiabatic steps, the irreversible quantum Carnot cycle possesses two relaxation strokes~\cite{quan07,quan14,xiao15}.

In this study, we investigate a single spin $1/2$ as the working medium of the irreversible quantum Carnot cycle. Recently, a nuclear spin $1/2$ have been implemented in the liquid state NMR setup as a quantum Otto heat engine operating at a finite-time~\cite{peterson19,assis19}. Here, we use the Hamiltonian driving protocols and the system parameters that are applicable to the liquid state NMR platform~\cite{peterson19,assis19,batalhao15,batalhao14}. Our results demonstrate that the quantum Carnot cycle using a nuclear spin $1/2$ is a promising candidate to be implemented in the liquid state NMR setup. We investigate in detail the effect of the time allocated to the adiabatic branches on the cycle efficiency and the work output. The performance outputs of the cycle are found to be dramatically influenced by the time length of the driving protocol. The Carnot efficiency is reached only at the quasi-static regime. The fast Hamiltonian driving protocol, however, reduces the work output and the efficiency. Moreover, it is shown that the quantum Carnot cycle can be incapable to produce useful work for the too fast adiabatic stages.

A single spin is inherently a scale invariant system. However, the inter-level transitions make the spin at non-equilibrium state at the end of the adiabatic stages. When the spin becomes in contact to the reservoir, an amount of heat is found to be released by the spin before the isothermal step. The amount of the released heat in the relaxation processes is connected both to the quantum friction arises in the adiabatic stages and to the entropy distance between the final states approached at the end of the adiabatic and the relaxation strokes. It is shown that the contributions on the released heat during the relaxation can be separated into two parts, one due to the coherence generation and the other one due to the incoherent transitions. The extension to the general quantum systems possessing scale invariance is also discussed.

\section{Irreversible quantum Carnot cycle}

\begin{figure}[!ht]\centering
\includegraphics[width=7.0cm]{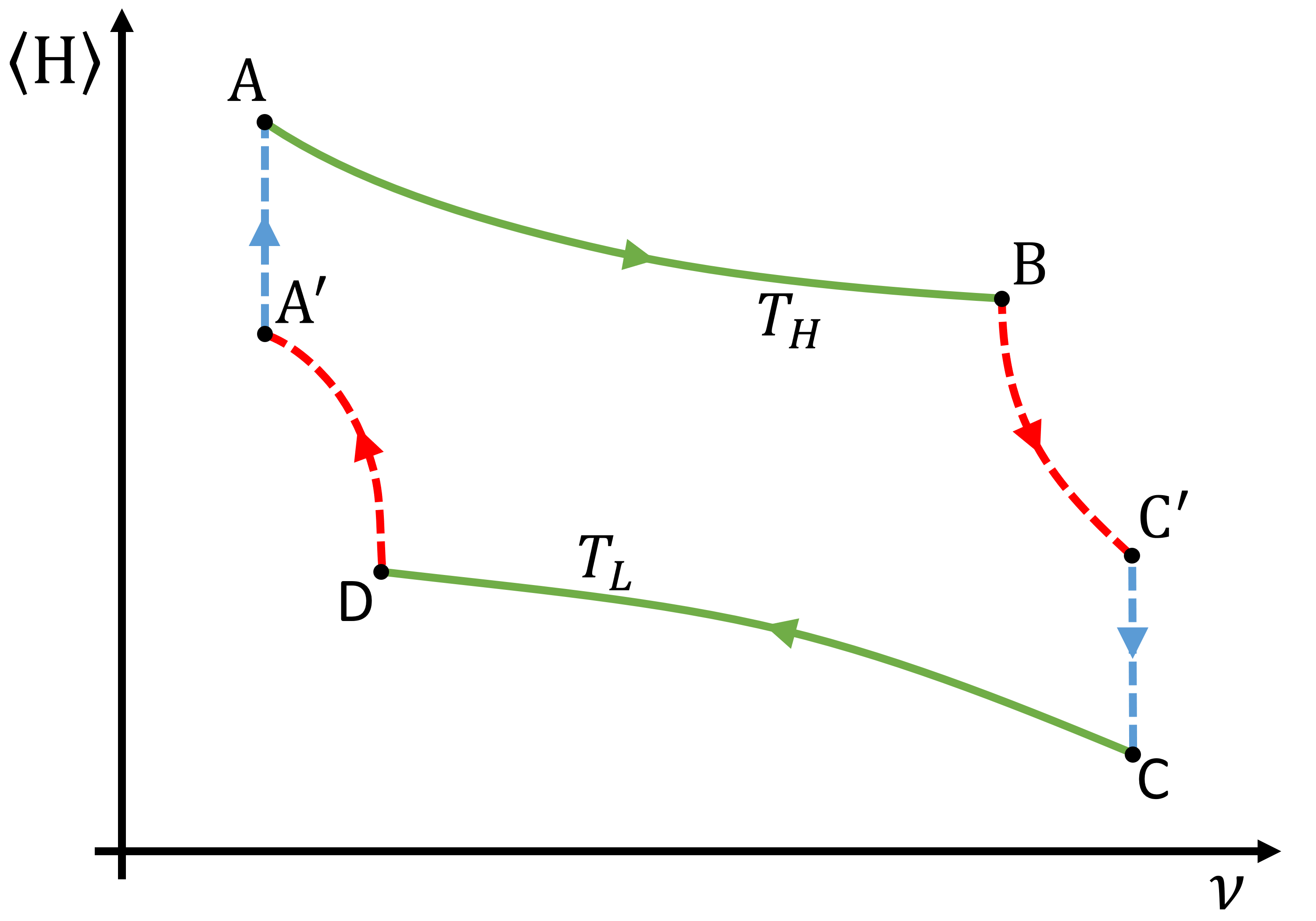}
\caption{\label{fig1} The schematic diagram of the six stroke irreversible quantum Carnot cycle. Here $\left\langle H \right\rangle$ is mean energy of the spin and $\nu$ is the tunable parameter in the cycle. $A \rightarrow B$ $\left(C \rightarrow D \right)$ is the quantum isothermal process where the spin is always at thermal equilibrium with the hot (cold) heat bath at temperature $T_H$ $\left( T_L \right)$. $B \rightarrow C^{'}$ $\left(D \rightarrow A^{'} \right)$ is the unitary process. The spin Hamiltonian is driven in a finite time interval which evolves the initial equilibrium state at instant $B$ $(D)$ to a non-equilibrium one at the end of the process $C^{'}$ $(A^{'})$. A relaxation with the cold (hot) reservoir occurs in the process $C^{'} \rightarrow C$  $\left( A^{'} \rightarrow A \right)$ at a fixed frequency $\nu_C$ $(\nu_A)$. }
\end{figure}

The six strokes of the irreversible quantum Carnot cycle can be simply schematized as in Fig.~\ref{fig1}. The details are given below.

(i) Isothermal heating stroke. The first stage $A \rightarrow B$ is the isothermal process. The Hamiltonian denoted as $H_A=-1/2h\nu_A\sigma_y$ at instant $A$ is changed to $H_B=-1/2h\nu_B\sigma_y$ at instant $B$ by only adjusting the field intensity. Here $h$ is the Planck constant and $\sigma_{x,y,z}$ are the Pauli operators. The process is slow enough to keep the spin always at thermal equilibrium with the hot reservoir at temperature $T_H$. The equilibrium spin state can be given as $\rho_{i}=1/Z_i e^{-\beta_H H_i}$ $(i=A,B)$, where $Z_i$ is the partition function, $\beta_H=1/(k_B T_H)$ and $k_B$ is the Boltzmann constant. The heat absorbed by the spin during the process can be written as $Q_{AB}=T_H\left(S_B-S_A\right)$, with $S_i=-k_B tr(\rho_i \ln\rho_i)$ being the thermodynamic entropy.

(ii) Adiabatic compression stroke. In the process $B \rightarrow C^{'}$, the system Hamiltonian is driven according to $H_{comp}(t)=-\frac{1}{2}h\nu(t)\left[\cos\left(\frac{\pi t}{2\tau}\right)\sigma_y+\sin\left(\frac{\pi t}{2\tau}\right)\sigma_x\right]$, with $\nu(t)=\nu_B(1-t/\tau)+\nu_C t/\tau$ and $\nu_C=\left(T_L/T_H\right)\nu_B$. The initial spin Hamiltonian $H_B=-1/2h\nu_B\sigma_y$ is driven to $H_{comp}(\tau)=-1/2h\nu_C\sigma_x \equiv H_C$ at time $t=\tau$. We remark here that the adopted driving protocol has been experimentally studied recently~\cite{peterson19,assis19,batalhao15,batalhao14}. The driving time length $\tau$ is much shorter than the decoherence time scales and the process is, therefore, unitary. The time development of the spin density matrix can be given as $\dot{\rho}(t)+i/\hbar\left[H_{comp}(t),\rho(t)\right]=0$. The spin state at the end of the process is denoted as $\rho_{C^{'}}$. Work is done on the spin.

(iii) Relaxation. Due to the finite time nature of the preceding stroke (ii), transitions among the instantaneous eigenstates of the spin Hamiltonian occur. An energy frame coherence may be generated which can make the spin state $\rho_{C^{'}}$ out of thermal equilibrium. When the spin becomes in contact to the cold heat bath at inverse temperature $\beta_L$, a relaxation process may occur. The state $\rho_{C^{'}}$ approaches to the thermal state $\rho_{C}=1/Z_C e^{-\beta_L H_C}$ at the end of the process $C^{'} \rightarrow C$. Since the frequency $\nu_C$ is considered unchanged during the relaxation, the stage can be interpreted as the quantum isochoric process. The heat exchanged during the relaxation can be given as $Q_{C^{'}C}=tr\left(H_C(\rho_C-\rho_{C^{'}})\right)$. As we will discuss later in the text that $Q_{C^{'}C}$ is always negative and can be interpreted as the quantum friction arises in the adiabatic compression stroke (ii).

(iv) Isothermal cooling stroke. The process $C \rightarrow D$ is almost inverse of the isothermal heating stroke (i). The spin Hamiltonian $H_C=-1/2h\nu_C\sigma_x$ at instant $C$ is slowly changed to $H_D=-1/2h\nu_D\sigma_x$ at instant $D$. The spin states at the terminal points are given as $\rho_{i}=1/Z_i e^{-\beta_L H_i}$ $(i=C,D)$. The released heat by the spin to the sink is calculated as $Q_{CD}=T_L\left(S_D-S_C\right)$.

(v) Adiabatic expansion stroke. In the process $D \rightarrow A^{'}$, the time dependent field drives the spin Hamiltonian according to $H_{exp}(t)=-\frac{1}{2}h\nu(t)\left[\cos\left(\frac{\pi t}{2\tau}\right)\sigma_x+\sin\left(\frac{\pi t}{2\tau}\right)\sigma_y\right]$, with $\nu(t)=\nu_D(1-t/\tau)+\nu_A t/\tau$ and $\nu_A=\left(T_H/T_L\right)\nu_D$. The initial Hamiltonian $H_D=-1/2h\nu_D\sigma_x$ is changed to $H_A=-1/2h\nu_A\sigma_y$ at the end. The process is unitary. The spin state denoted as $\rho_{A^{'}}$ at time $t=\tau$ is, in general, not an effective thermal state.

(vi) Relaxation. In the process $A^{'} \rightarrow A$, the non-equilibrium state $\rho_{A^{'}}$ reaches to the thermal state $\rho_{A}=1/Z_A e^{-\beta_H H_A}$ by exchanging heat with the hot reservoir at temperature $T_H$. As we will show, the mean energy at the instant $A^{'}$ is always greater than the one at $A$. Therefore, the exchanged heat $Q_{A^{'}A}=tr\left(H_A(\rho_A-\rho_{A^{'}})\right)$ is always negative.

The net heat dumped to the sink (defined as $Q_{out}$) and the net heat absorbed from the source (defined as $Q_{in}$) are, respectively, given as
\begin{eqnarray}\label{heat}
Q_{out}&=&Q_{C^{'}C}+Q_{CD}\nonumber\\
&=&tr\left(H_C(\rho_C-\rho_{C^{'}})\right)+T_L\left(S_D-S_C\right)\nonumber\\
Q_{in}&=&Q_{A^{'}A}+Q_{AB}\nonumber\\
&=&tr\left(H_A(\rho_A-\rho_{A^{'}})\right)+T_H\left(S_B-S_A\right).
\end{eqnarray}
The amount of net work $W=Q_{in}+Q_{out}$ is extracted from the irreversible Carnot cycle when $Q_{in}>-Q_{out}>0$. The heat-to-work efficiency is then defined as
\begin{eqnarray}\label{efficiency}
\eta=1+\frac{Q_{out}}{Q_{in}}.
\end{eqnarray}

\section{Results and Discussion}
As a next step, we give the numerical values of the parameters defined above. Here we use the frequencies and the bath temperatures which are applicable to the liquid state NMR setup~\cite{peterson19,assis19,batalhao15,batalhao14}. The fixed parameters are $\nu_D=2.0$ kHz, $\nu_B=3.6$ kHz and $T_L=6.6$ peV/$k_B$. Different source temperatures are considered $T_H=(16.5,21.5,26.5,33.5)$ peV/$k_B$. We should stress here that as the temperature ratio $T_H/T_L$ changes, the frequencies at the instants $A$ and $C$ also change. However, they are still few keVs and practical for NMR platform.

\begin{figure}[!ht]\centering
\includegraphics[scale=0.205]{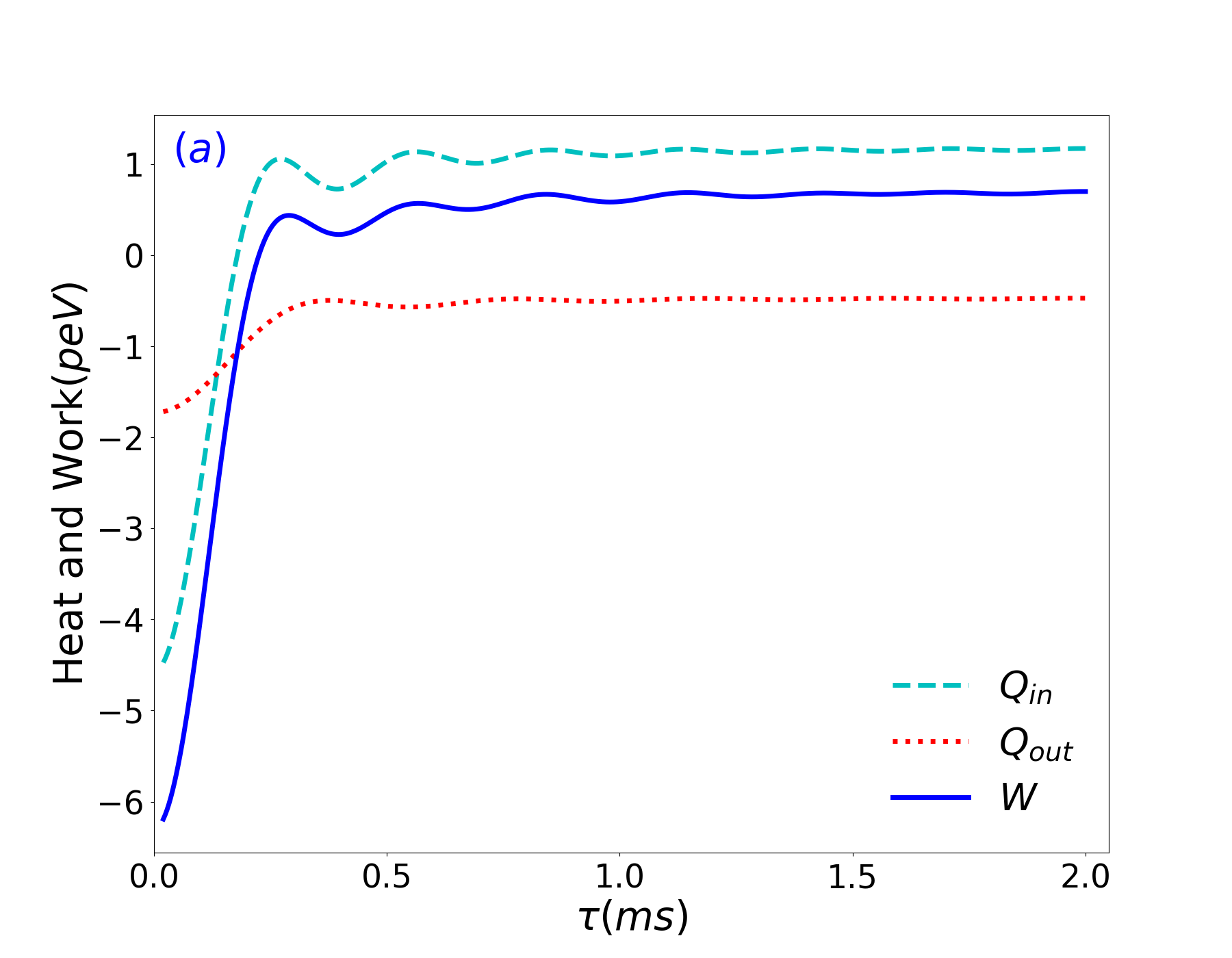}
\includegraphics[scale=0.205]{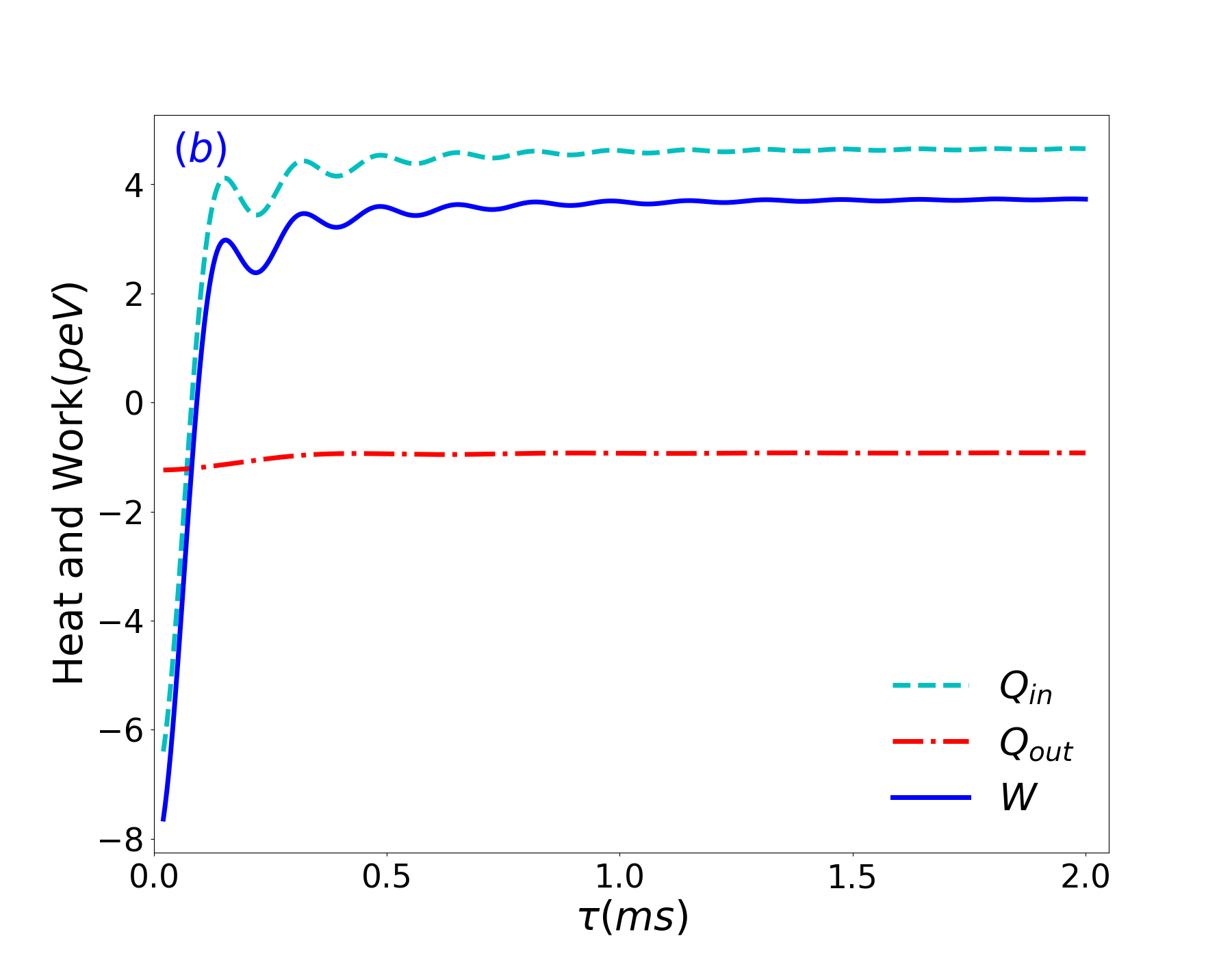}
\caption{\label{fig2} Average work (solid line) and the net heat exchanged with the hot (dashed line) and the cold (dotted line) reservoirs versus the driving time length, $\tau$. Here we set the source temperature at $T_H=16.5$ peV/$k_B$ in (a) and at $T_H=33.5$ peV/$k_B$ in (b).}
\end{figure}

In Fig.~\ref{fig2}, we plot the average work and heat per cycle versus the driving time length for the source temperature $T_H=16.5$ peV/$k_B$ (Fig.~\ref{fig2}(a)) and $T_H=33.5$ peV/$k_B$ (Fig.~\ref{fig2}(b)). Figure illustrates that the exchanged heat and work are dramatically influenced by the time length of the driving protocol. Slower operation makes the spin to harvest more positive work. However, when the time length ($\tau$) decreases, the extracted work and the absorbed heat from the source decrease almost monotonically. In addition, the fast driving protocol makes the spin to dump more heat to the cold sink. This shows that too fast driving protocols inhibit the heat engine operation, indicating the presence of dissipation due to the irreversibility. Specifically, as shown in Fig.~\ref{fig2}(a), work is injected to the spin if $\tau$ is approximately smaller than $220$ $\mu s$. For the driving times $\tau < 220$ $\mu s$, the operation mode of the cycle is not a refrigerator; work and heat are all negative in this regime, so the operation of the cycle has no industrial use. On the other hand, as the temperature of the hot bath is set at $T_H=33.5$ peV/$k_B$ (Fig.~\ref{fig2}(b)), the spin can harvest positive work even for a too fast driving protocol (i.e, $\tau < 220$ $\mu s$); the spin becomes incapable to produce useful work only for the driving times smaller than $90$ $\mu s$. Increasing the temperature of the source bath increases the magnitude of the harvested work due to the capability of the spin to gain more heat from the source.

\begin{figure}[!ht]\centering
\includegraphics[scale=0.205]{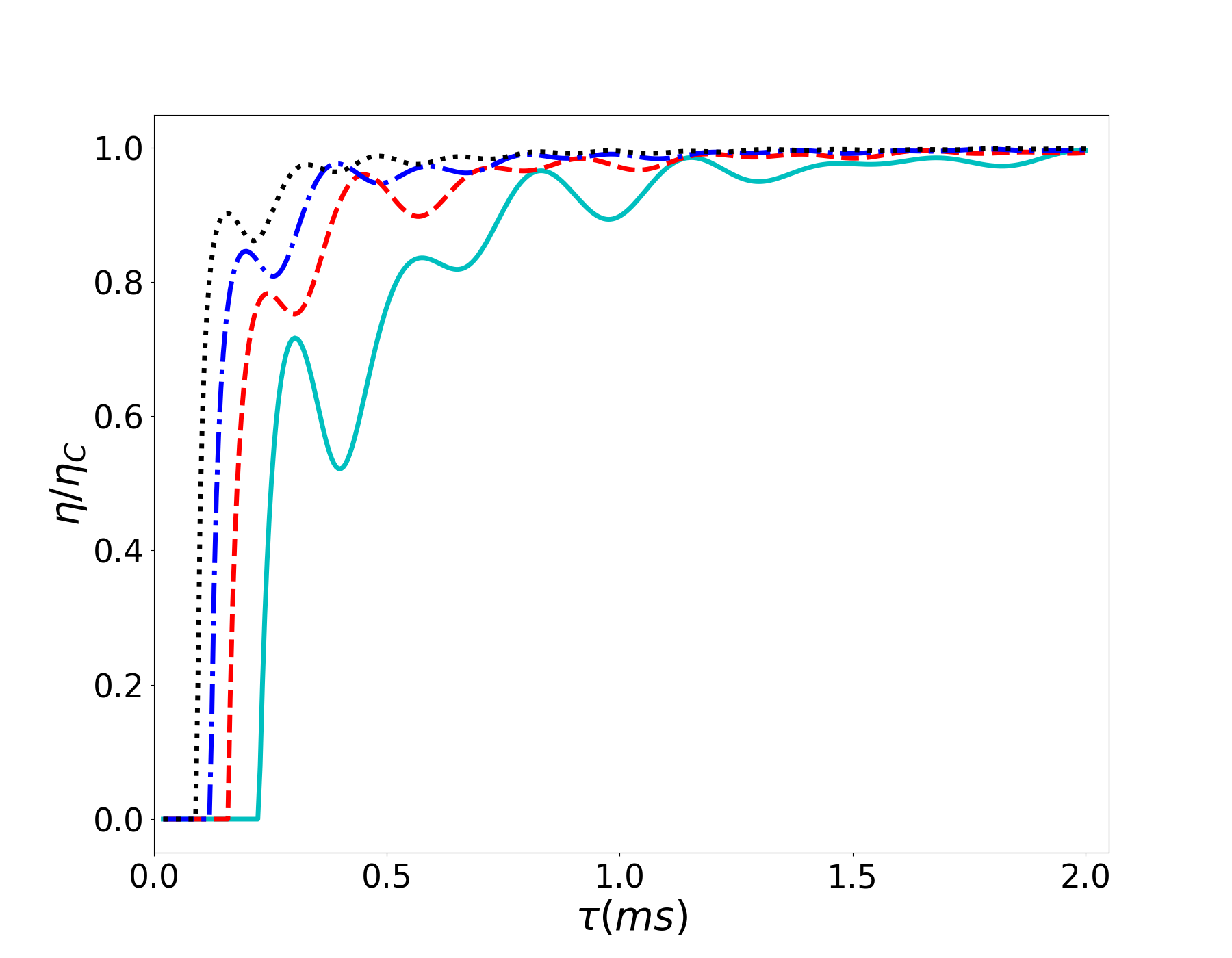}
\caption{\label{fig3} The efficiency of the six stroke irreversible quantum Carnot cycle (Eq.~(\ref{efficiency})) divided by the classical Carnot efficiency ($\eta_C=1-\beta_H/\beta_L$), i.e., $\eta/\eta_C$ versus the driving time length, $\tau$. Here the lines labeled as (solid, dashed, dot-dashed, dotted) are for $T_H=(16.5,21.5,26.5,33.5)$ peV/$k_B$, respectively.}
\end{figure}

Now we investigate the effect of the irreversibility associated with the finite time Hamiltonian driving on the cycle efficiency. In Fig.~\ref{fig3}, we plot the efficiency (Eq.~(\ref{efficiency})) divided by the classical Carnot efficiency ($\eta_C=1-\beta_H/\beta_L$) versus $\tau$ for the source temperatures $T_H=(16.5,21.5,26.5,33.5)$ peV/$k_B$. Fast adiabatic operations are also worse for the cycle efficiency. The ratio $\eta/\eta_C$ decays to zero as $\tau$ decreases and the heat engine character disappears after a critical driving time. As shown from the figure, the critical driving time is sensitive to the hot bath temperature. When $T_H/T_L$ ratio increases, the critical driving time shifts to the smaller values. In addition, higher source temperature results in more efficient cycle for a given time $\tau$. We may claim here that the cycle efficiency is more effected by the irreversibility for a small source temperature compared to the higher one. On the other hand, when the driving protocol time increases, the ratio $\eta/\eta_C$ expeditiously increases and approaches to 1 ($\eta/\eta_C \rightarrow 1$) as $\tau \rightarrow \infty$ for all temperature cases. We remark here that for the intermediate driving time intervals, the efficiency and also the work and heat have non-monotone dependence on the protocol time length $\tau$. As we will discuss later in detail that the quantum coherence generation due to the driving is responsible to such an unexpected dependence.

Once the average work and heat in Fig.~\ref{fig2} and the efficiency in Fig.~\ref{fig3} saturate as a function of $\tau$, they do no longer change for the further increase of the driving time length~\cite{allah05}. The quantum adiabatic results are obtained in which the total entropy increase of the universe is zero. The quantum Carnot cycle is reversible and we obtain $\eta/\eta_C \rightarrow 1$ in the limit $\tau \rightarrow \infty$. On the other hand, the deviation from the quantum adiabatic results is the signature of the irreversibility in the quantum Carnot cycle. The effect is fully a quantum phenomenon and called as the inner friction~\cite{thomas14,allece15,rezek06,kosloff02,feldmann06,plastina14,deffner10,francica19,camati19,rezek10,kosloff13}. The irreversibility is the consequences of the time ordering in the unitary time evolution which arises when the driving Hamiltonians at different time instants are incompatible, i.e., $\left[H(t_1),H(t_2)\right] \neq 0$.

The inner friction in the above proposed Carnot cycle can be directly characterized by the entropy distance between the spin state reached after the adiabatic strokes (i.e., $\rho_{C^{'}}$ and $\rho_{A^{'}}$) and the thermal states at the beginning  of the isothermal stages (i.e., $\rho_{C}$ and $\rho_{A}$). This entropy distance is intimately related to the heat exchanged during the relaxation processes of the cycle (i.e., $Q_{C^{'}C}$ and $Q_{A^{'}A}$).

Let us first concentrate the Hamiltonian driving protocol $B \rightarrow C^{'}$. At instant $B$, the spin is in a thermal state $\rho_B=\sum_n p_n^B \left|n_B\right\rangle\left\langle n_B\right|$ with the Hamiltonian $H_B=\sum_n \epsilon_n^B\left|n_B\right\rangle\left\langle n_B\right|$. Here and in the following, $\epsilon_n^i$ ($\left|n_i\right\rangle$) denote the energies (the eigenfunctions) and $p_n^i$ are the thermal Boltzmann weights. In the adiabatic compression stroke, the spin Hamiltonian is driven from $H_B$ to $H_C=\sum_n \epsilon_n^C\left|n_C\right\rangle\left\langle n_C\right|$. If the process is quasistatic (i.e., $\tau \rightarrow \infty$), the spin state follows the instantaneous changes in the energies without changing the initial level populations. Therefore, the spin state at the end would be $\rho_{C^{'}}=\sum_n p_n^B \left|n_C\right\rangle\left\langle n_C\right|$ as predicted by the quantum adiabatic theorem. The spin state at the instant $C^{'}$ is infact a thermal state with an effective temperature $T_{C^{'}}$. The spin state can be rewritten as $\rho_{C^{'}}=\sum_n p_n^{C^{'}} \left|n_C\right\rangle\left\langle n_C\right|$. By getting the ratio $p_n^{C^{'}}/p_m^{C^{'}}=p_n^{B}/p_m^{B}$ and using the predefined frequency $\nu_C=(T_L/T_H)\nu_B$, one can easily prove that $T_{C^{'}} \equiv T_L$. In other words, for a quasistatic transformation, the spin state at the instant $C^{'}$ would be $\rho_{C^{'}}\equiv\rho_{C}$.  When the spin becomes in contact to the cold bath, there would be no relaxation. A similar analysis for the adiabatic expansion stroke $D \rightarrow A^{'}$ gives $\rho_{A^{'}}\equiv\rho_{A}$, where $Q_{A^{'}A}=0$ in the limit  $\tau \rightarrow \infty$. Since the instants $A^{'}$ and $C^{'}$ coincide with the instants $A$ and $C$ in the limit $\tau \rightarrow \infty$, the efficiency in Eq.~(\ref{efficiency}) exactly gives the classical Carnot efficiency. The above analysis indeed explains why the ratio $\eta/\eta_C$ in Fig.~\ref{fig3} goes to 1 as the parameter $\tau$ increases.

For a finite time Hamiltonian driving, the final spin states $\rho_{C^{'}}$ and $\rho_{A^{'}}$  are out of equilibrium due to the inter-level transitions. One can investigate the closeness of the non-equilibrium spin states to their equilibrium correspondences through quantum relative entropy, i.e., $S(\rho_{i^{'}}||\rho_{i})$ ($i=A,C$), where $S(\rho||\sigma)=tr(\rho(\ln\rho-\ln\sigma))$. The relative entropy defined just above has a well thermodynamic interpretation~\cite{plastina14}. It is related to the inner friction arises in the adiabatic strokes and also to the heat exchanges during the relaxation processes of the cycle.

The inner friction ($W_{fric}$) arises from the finite time Hamiltonian driving is defined as the work done on an actual finite time process minus the work done on a quasistatic transformation~\cite{plastina14}. Since the spin states under quasistatic transformation are exactly the thermal states at the beginning of the isothermal stages, we arrive the following relations $W_{fric}^{B \rightarrow C^{'}}=tr\left(H_C(\rho_{C^{'}}-\rho_{C})\right)=-Q_{C^{'}C}$ and $W_{fric}^{D \rightarrow A^{'}}=tr\left(H_A(\rho_{A^{'}}-\rho_{A})\right)=-Q_{A^{'}A}$. Using the spectral decomposition of the relevant density matrices and the invariant property of the von Neumann entropy in a unitary transformation, the above defined relative entropy can be written in a form as~\cite{plastina14,deffner10} $S(\rho_{i^{'}}||\rho_{i})=\beta_i tr\left(H_i(\rho_{i^{'}}-\rho_{i})\right)$ ($i=A,C$). As a result, we arrive the following equation
\begin{eqnarray}\label{wfric}
W_{fric}^{B \rightarrow C^{'}}&=&\beta_L^{-1}S(\rho_{C^{'}}||\rho_{C})=-Q_{C^{'}C},\nonumber\\
W_{fric}^{D \rightarrow A^{'}}&=&\beta_H^{-1}S(\rho_{A^{'}}||\rho_{A})=-Q_{A^{'}A}.
\end{eqnarray}
$S(\rho||\sigma)$ is always non-negative~\cite{plastina14}. Therefore, the heat exchanges during the relaxation processes are always negative, $Q_{A^{'}A},Q_{C^{'}C}<0$. This can be understood as follows. The driving agent should supply more energy on the spin to make the same changes in a fast driving protocol. This result in extra energy to be stored in the spin as quantified by $W_{fric}$. The spin density matrix deviates from equilibrium and becomes a non-equilibrium state. To approach the adiabatically evolved equilibrium state defined at the beginning of the isothermal stages, the extra stored energy is released to the heat baths in the relaxation processes of the cycle. Therefore, the finite-time adiabatic processes and the following inevitable relaxation strokes make the quantum Carnot cycle irreversible, which decreases the positive work output and the operational efficiency due to the increase (decrease) the heat released (absorbed) to (from) the cold (hot) heat reservoir.

The irreversibility associated with the finite-time Hamiltonian driving introduces the engine efficiency a departure to the ideal Carnot efficiency, $\eta_C=1-\beta_H/\beta_L$ (see Fig.~\ref{fig3}). We can find an explicit expression for the efficiency (Eq.~(\ref{efficiency})) in terms of the efficiency lag~\cite{peterson19,camati19, campisi16}, i.e, $\eta=\eta_C-\mathcal{L}$. Using Eqs.~(\ref{heat}),~(\ref{efficiency}) and~(\ref{wfric}), and the invariant property of the von Neumann entropy in a unitary transformation, the lag can be found after a simple calculation as
\begin{eqnarray}\label{efflag}
\mathcal{L}=\frac{S(\rho_C^{\prime} || \rho_C) + S(\rho_A^{\prime} || \rho_A)}{\beta_L Q_{in}}.
\end{eqnarray}
The lag in Eq.~(\ref{efflag}) is in the same form as the one recently given for a finite-time Otto cycle ~\cite{peterson19,camati19}. The lag $\mathcal{L}$ quantifies the detrimental contribution of the quantum friction on the cycle efficiency. For a finite-time driving, $\mathcal{L}$ is always positive so $\eta<\eta_C$. The efficiency lag becomes zero only in the absence of inner friction where $\eta=\eta_C$.

The contributions on the inner friction can be divided into two parts~\cite{francica19,camati19}. The first is coming from the energy frame coherence generated in the non-equilibrium state $\rho_{i^{'}}$. The second contribution is coming from the population difference between the non-equilibrium state $\rho_{i^{'}}$ and the equilibrium correspondence $\rho_{i}$. A recent study shows to quantify the amount of the separate contributions as~\cite{francica19}
\begin{eqnarray}\label{separation}
S(\rho_{i^{'}}||\rho_{i})=C(\rho_{i^{'}})+S(\Delta(\rho_{i^{'}})||\rho_i),
\end{eqnarray}
where $i=A,C,$ and $\Delta(\rho_{i^{'}})$ is the energy frame dephasing map. The first term on the right hand side measures the amount of coherence contribution, while the second term quantifies the amount of the population mismatch contribution. The detailed calculation procedure of the contributions in Eq.~(\ref{separation}) can be found in Ref.~\cite{francica19}.

\begin{figure}[!ht]\centering
\includegraphics[scale=0.205]{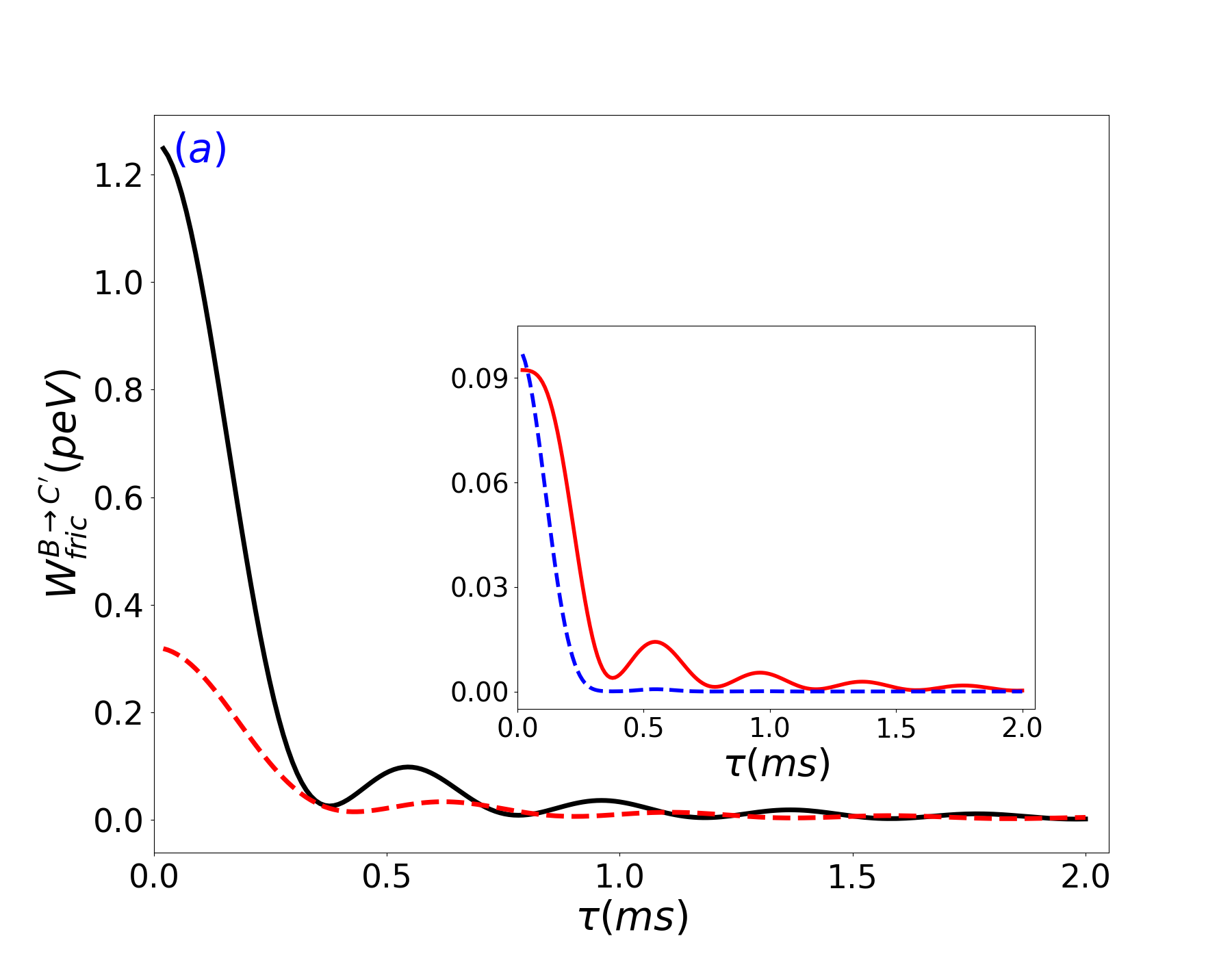}
\includegraphics[scale=0.205]{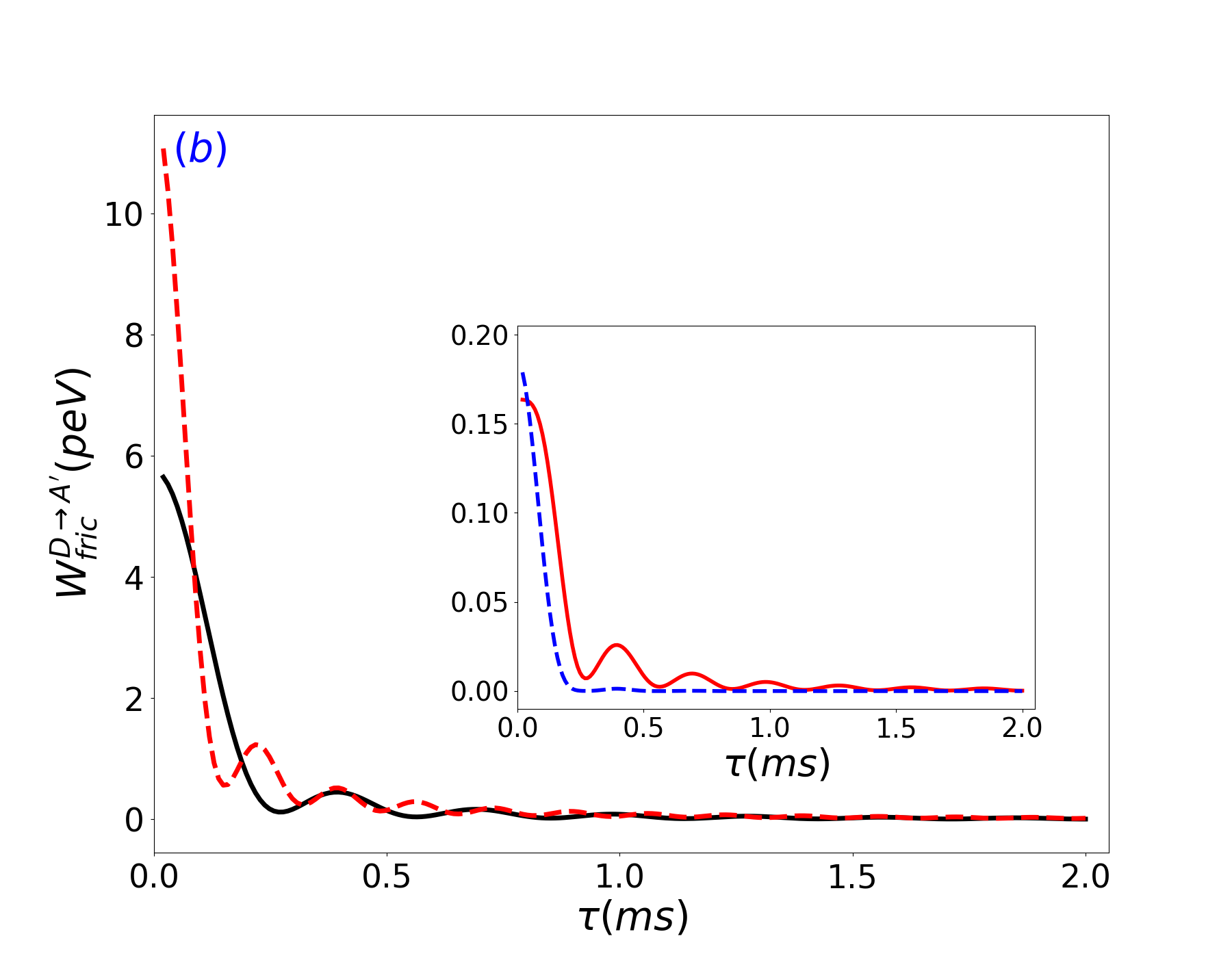}
\caption{\label{fig4} The inner friction arises in the adiabatic compression (a) and expansion (b) strokes versus the driving time length, $\tau$ for the source temperatures $T_H=16.5$ peV/$k_B$ (solid line) and $T_H=33.5$ peV/$k_B$ (dashed line). The inner friction here coincides with the heat dumped to the reservoirs during the relaxation processes. The insets show the contribution parts to the nonadiabaticity by the coherence generation (solid line) and by the population mismatch (dashed line) versus $\tau$ for the source temperature $T_H=16.5$ peV/$k_B$.}
\end{figure}

In Fig.~\ref{fig4}, we plot the inner friction (Eq.~(\ref{wfric})) arises from the fast Hamiltonian driving and the contribution parts on the inner friction (Eq.~(\ref{separation})) as a function of the protocol time length ($\tau$). Please remark here again that the inner friction in the sub-figures is exactly equal to the heat dumped to the reservoirs during the relaxation processes  (Eq.~(\ref{wfric})). The inner friction is high for a too fast Hamiltonian driving protocol which decreases almost monotonically and vanishes when $\tau$ is large enough. As shown from the insets, both coherence generation and population mismatch contributions have significant effects for the fast driving case. As $\tau$ increases, the contributions become less and vanish in the limit $\tau\rightarrow\infty$. $C(\rho_{i^{'}})$ in Eq.~(\ref{separation}) exactly matches the production of Shannon entropy defined in terms of the energy level populations~\cite{francica19}. Therefore, the coherence contribution to the nonadiabaticity indicates the production of entropy due to the inter-level transitions. The common expectation is the monotonic decay of $W_{fric}$ as a function of $\tau$. However, as shown from Fig.~\ref{fig4} and also from the Figs.~\ref{fig2} and~\ref{fig3}, the figure of merits for the intermediate time intervals have a non-monotone dependence on $\tau$. This is not just a theoretical prediction obtained from a unitary time evolution which is recently reported experimentally in a liquid state NMR setup~\cite{peterson19,assis19,batalhao15,batalhao14}. The insets illustrate that $C(\rho_{i^{'}})$ gives the leading contribution both on the non-monotonic feature and the nonadiabaticity for the slow driving case. The population mismatch contribution here decays quickly as $\tau$ increases.

Eqs.~(\ref{wfric}) and~(\ref{efflag}) also hold for a general working substance if two conditions are satisfied: (1) all energy gaps are altered by the same ratios in the two adiabatic steps, i.e., $\epsilon_n^B-\epsilon_m^B=\lambda \left(\epsilon_n^C-\epsilon_m^C\right)$ and $\epsilon_n^A-\epsilon_m^A=\lambda \left(\epsilon_n^D-\epsilon_m^D\right)$ and (2) the proportionality ratio $\lambda$ is equal to the ratio of the two bath temperatures, i.e., $\lambda=T_H/T_L$. For a quasistatic adiabatic transformation, the first condition guarantees the thermal state, while the latter one ensures the effective temperature being equal to the bath temperature. In other words, the states $\rho_{C^{'}}$ and $\rho_{A^{'}}$ coincide with $\rho_{C}$ and $\rho_{A}$ under the conditions (1) and (2) in the quasistatic limit. Then the quasistatic Carnot engine is thermodynamically reversible~\cite{quan07}. Consider now a time dependent Hamiltonian $H(t)$ generates the adiabatic strokes of the cycle. If the Hamiltonian is incompatible at different times, a finite time driving makes the density matrices out of the thermal equilibrium. The thermalization of the non-equilibrium state with the reservoir then occurs before the isothermal process. More precisely, the parasitic internal energy stored in the driving protocol is released in the relaxation stroke as a heat which are expressed as the quantum relative entropy between the density matrices at the end of the adiabatic and relaxation steps (Eq.~(\ref{wfric})). The departure of the cycle efficiency to the classical Carnot efficiency due to the quantum friction is then given by Eq.~(\ref{efflag}) for a general working substance under the conditions (1) and (2).

As a final remark we would like to discuss the possibility to implement the irreversible quantum Carnot cycle in a liquid state NMR setup~\cite{peterson19,assis19,batalhao15,batalhao14}. Recently, a proof-of-concept realization of a quantum Otto cycle has been given in a liquid state NMR platform~\cite{peterson19,assis19}. The spin $1/2$ of $^{13}$C nucleus as a working substance encounters two quantum isochoric steps and two Hamiltonian driving protocols. Due to the fast thermal-relaxation time of $^{1}$H nucleus in contrast to $^{13}$C, $^{1}$H can thermalize the $^{13}$C nucleus via $J$-coupling effect. Therefote, the spin $1/2$ of $^{13}$C nucleus is coupled to the spin $1/2$ of $^{1}$H nucleus in order to simulate the interaction of the working medium with the hot heat bath during the isochoric stage. $^{1}$H and $^{13}$C have different gyromagnetic ratio, thus it is possible to manipulate each nucleus seperately using NMR selective pulses. Transverse radio frequency pulses are used to prepare the nuclei in desired effective thermal states. The thermal states are long lived due to the long thermal relaxation times in NMR.  In the adiabatic stages, the $^{13}$C nuclear spin is driven by a resonant time-dependent radio frequency field. The adiabatic steps are completed in few milliseconds. The characteristic decoherence time scales of $^{13}$C nuclear spin is in order of seconds. Therefore, the time development of the driving protocols can be considered as a unitary dynamics. The density matrices at the terminal points are determined by the tomography of states. The work and efficiency are obtained experimentally and also predicted theoretically. It was shown that the experimental outputs are in full accordance with the theoretical predictions. In addition to two isochoric and two adiabatic steps, the irreversible quantum Carnot cycle given in Fig.~\ref{fig1} possesses two quantum isothermal stages. The isothermal stages can be realized by putting the working substance in contact to a reservoir while changing only the field intensity. The speed of the change should be slow to keep the working medium always at thermal equilibrium with the heat bath. Our results demonstrate that highly qualified quantum Carnot heat engine can be obtained for the driving time length smaller than $1.5$ ms. This time is much shorter than the decoherence times in NMR. Therefore, we may claim here that the proposed irreversible quantum Carnot cycle is a promising candidate to be implemented in a liquid state NMR platform.

\section{Conclusions}
In this paper, we have investigated a driven spin as the working medium of an irreversible quantum Carnot cycle. The role of the finite time adiabatic operations on the cycle performance has been investigated in detail. The finite time operations are the source of the irreversibility named as the inner friction.  The irreversibility dramatically reduces the work output and the cycle efficiency which can make the engine incapable to produce  work for the too fast adiabatic transformations. The ideal Carnot efficiency is reached only  at the quasi-static regime.

The driving agent should supply more energy (as quantified by inner friction) on the spin to make the same changes for a fast driving protocol. The extra energy should be released to the heat bath before the isothermal processes of the cycle. A compact relationship has been given between these quantities and the quantum relative entropy of the density matrices obtained at the end of the adiabatic and the relaxation steps. A deviation of the cycle efficiency from the ideal Carnot efficiency due to the quantum friction has been represented by an efficiency lag which is intimately related to the total entropy production due to the quantum friction.  The contributions on the inner friction  can be separated into two parts, one due to the coherence generation and the other one due to the incoherent transitions. It has been found that  the coherence generation gives the leading contribution both on the non-monotonic feature as a function of the driving time length and the nonadiabaticity for the slow driving transformation. We have been noted that the results are also valid for a general quantum system possessing scale invariance. We believe that our results are eligible to the experimental implementation, especially in a liquid state NMR platform.


\end{document}